\documentstyle[preprint,aps]{revtex}
\def\go{                                                                                                        
\mathrel{\raise.3ex\hbox{$>$}\mkern-14mu\lower0.6ex\hbox{$\sim$}}                                               
}                                                                                                               
\def\lo{                                                                                                        
\mathrel{\raise.3ex\hbox{$<$}\mkern-14mu\lower0.6ex\hbox{$\sim$}}                                               
} 
\begin{document}

\title{
Numerical Integration of Nonlinear Wave
Equations for General Relativity}

\author{Maurice H.P.M. van Putten 
	\footnote
        {Electronic address: 
	\tt mvp@schauder.mit.edu\hfil}\\
CRSR \& Cornell Theory Center,
        Cornell University,
        Ithaca, NY 14853-6801, and Department of Mathematics, MIT,
	Cambridge, MA 02139.}

\date{\today}


\maketitle
\mbox{}\\

\begin{abstract}
A second-order numerical implementation is given for 
recently derived nonlinear wave equations  
for general relativity.
The Gowdy T$^3$ cosmology is used as a test bed for 
studying the accuracy and convergence of simulations of
one-dimensional 
nonlinear waves.
The complete freedom in space-time slicing in the 
present formalism is
exploited to compute in the Gowdy line-element.
Second-order convergence 
is found by direct comparison of the results
with either analytical solutions for polarized waves,
or solutions obtained from Gowdy's reduced
wave equations for the more general 
unpolarized waves. Some directions for extensions
are discussed.\\
\mbox{}\\
PACS numbers: 04.25.Dm, 04.30.Nk
\end{abstract}
\bibliographystyle{plain}
\section{Introduction}

The gravitational wave observatories LIGO and VIRGO
\cite{abramovici:a,bradaschia:a}
presently under construction have given added impetus towards
accurate prediction of gravitational wave forms
from binary coalescence of neutron stars and black holes.
The late stages of the spiral infall are predominantly targeted
through large scale simulations in numerical relativity.
The gravitational wave interactions
have motivated a description of gravity by nonlinear wave equations
in a tetrad approach (\cite{mvp:IX}). 
In this formulation, the strictly hyperbolic
nature of the equations is independent of the
particular choice of foliation. The foliation is governed by
four tetrad lapse functions which are algebraically related
to the more familiar Hamiltonian
lapse and shift functions.

In this paper, a numerical implementation of the nonlinear
wave equations is given, and the performance is studied in
the Gowdy T$^3$ cosmology.
The Gowdy T$^3$ is a good test bed for one-dimensional nonlinear
wave motion of both polarized and unpolarized waves.
The numerical scheme is one-dimensional,
second-order in time and spectrally
accurate in space. 
The SO$(3,1,{\bf R})$-connections are evolved  
by implementation of the four-divergence of the Riemann tensor
and the Lorentz gauge on the connections. 
The tetrad elements are evolved by the equations of structure,
and treated as a system of 
ordinary differential equations in which the
fundamental matrix is a finite Lorentz transformation.
The Gowdy T$^3$ line-element
has been incorporated in the numerical scheme, so as to enable
direct comparison of the computed solution with either analytical
or numerical solutions obtained from Gowdy's reduced wave equations.
Convergence results are presented for both
polarized and unpolarized Gowdy waves.

The nonlinear wave equations for the connections,
$\omega_{a\mu\nu}$, of the tetrad 
elements, $\{(e_\mu)_a\}$
satisfy nonlinear wave 
equations of the Yang-Mills type.
They follow from a Lorentz gauge on the connections.
In vacuo,
they simplify to
\begin{eqnarray}
\hat{\Box}\omega_{a\mu\nu}-[\omega^c,\nabla_a\omega_c]_{\mu\nu}=0.
\label{nl}
\end{eqnarray}
Here, $\hat{\Box}=\hat{\nabla}^c\hat{\nabla}_c$ with $\hat{\nabla}_a$
the SO(3,1,${\bf R}$)-gauge covariant derivative satisfying
$\hat{\nabla}_a(e_\mu)_b
=\nabla_a(e_\mu)_b
+\omega_{a\mu}^{\hskip.1in\gamma}(e_\gamma)_b=0$.
The tetrad elements 
define the metric by $g_{ab}=(e_\mu)_a(e^\mu)_b$.
Here, contraction over the Greek indices is 
through $\eta_{\mu\nu}=
\mbox{diag}(-1,1,1,1)$.
Initial data in this second-order formalism includes
the initial data for
the familiar first-order Hamiltonian formalism.
Additional initial data are for example
initial values for the tetrad; with sufficient smoothness
these data generate initial data for the connections and 
the Riemann tensor.
A procedure for obtaining smooth initial tetrad data from 
the metric is included.

In Section 2, the Gowdy waves and their reduced wave equations are discussed.
Section 3 presents the integration scheme, together with the procedure
for computing initial data for the tetrad elements.
Section 4 discusses the
gauge conditions in the Gowdy line-element. Discussion of the simulations
and conclusions are given in Section 5.

\section{Gowdy waves}

Gowdy cosmologies describe an extensively studied
class of universes with compact space-like
hypersurfaces with two 
Killing vectors,
$\partial_\sigma$ and $\partial_\delta$.
By choice of boundary conditions, the space-like
hypersurfaces are homeomorphic to either
the three-torus, the three-handle or the three-sphere.
The three-torus describes a semi-infinite
time-evolution of universes collapsing into,
or beginning with
a singularity. 

The Gowdy three-torus cosmology has recently been
investigated numerically by Berger and
Moncrief \cite{bv:a}, and can be given in the line-element
\begin{eqnarray}
ds^2=&e^{-\frac{\lambda}{2}}e^{\frac{\tau}{2}}
(-e^{-2\tau}\mbox{d}\tau^2+\mbox{d}\theta^2)+
\mbox{d}\Sigma^2,
\label{gmetr}
\end{eqnarray}
where 
\begin{eqnarray}
\mbox{d}\Sigma^2=
e^{-\tau}[e^P\mbox{d}\sigma^2
+2e^PQ\mbox{d}\sigma\mbox{d}\delta
+(e^PQ^2+e^{-P})\mbox{d}\delta^2].
\end{eqnarray}
Here $\lambda=\lambda(\tau,\theta),P=P(\tau,\theta)$
and $Q=Q(\tau,\theta)$, by invariance
in the angular
coordinates 
$\sigma$ and $\delta$. 

In small amplitude waves, $e^{-\tau}P$ and $e^{-\tau}Q$ are the
amplitudes of the $\epsilon_+$ and $\epsilon_\times$ 
gravitational wave polarization tensors (Eqn. 3.17 in \cite{bv:a}).
The quantities $P$ and $Q$ satisfy Gowdy's reduced wave equations, 
which may be given as \cite{bv:a}:
\begin{eqnarray}
\begin{array}{rl}
Q_{\tau\tau}&=e^{-2\tau}Q_{\theta\theta}-2(P_\tau Q_\tau-e^{-2\tau}
	       P_\theta Q_\theta),\\
P_{\tau\tau}&=e^{-2\tau}P_{\theta\theta}
+e^{-2P}(Q_{\tau}^2-e^{-2\tau}Q^2_{\theta}).
\end{array}
\label{gwave}
\end{eqnarray}

The $\theta-$momentum and Hamiltonian constraints, 
respectively, are given by
\begin{eqnarray}
\begin{array}{rl}
\lambda_\theta&=
2(P_\theta P_\tau
+e^{2P}Q_\theta Q_\tau),\\
\lambda_\tau&=
[P_\tau^2+e^{-2\tau}P_\theta^2
+e^{2P}(Q_\tau^2+e^{-2\tau}
Q_\theta^2)].
\end{array}
\label{gcon}
\end{eqnarray}
Notice that the wave equations 
(\ref{gwave}) evolve in an unconstraint 
manner with respect to
(\ref{gcon}). 

These equations serve two purposes. They are used to generate initial
data, and can be readily integrated for obtaining reference solutions
against which the simulations 
can be compared.
A simplectic integration scheme has been given by \cite{bv:a}. 
For reasons of convenience,
a leap frog integration scheme in combination 
with spectrally accurate spatial differentiation by the Fast
Fourier Transform has been used here.
Thus, general Gowdy waves are at hand with high accuracy
as a reference in studying 
the simulations in the tetrad approach.
In the polarized case, furthermore, analytical
solutions are available.

\subsection{A polarized Gowdy wave}

Polarized Gowdy wave result in the special case $Q=0$, so that
the Gowdy line-element becomes
\begin{eqnarray}
ds^2=e^{\frac{\tau-\lambda}{2}}
(-e^{-2\tau}\mbox{d}\tau^2+\mbox{d}\theta^2)
+e^{-\tau}[e^P\mbox{d}\sigma^2+
e^{-P}\mbox{d}\delta^2].
\label{gmetr0}
\end{eqnarray}
The reduced wave equation (\ref{gwave}) is linear in $P$, 
and can be solved by the method of
separation of variables, giving 
\cite{bv:a}
\begin{eqnarray}
P(\tau,\theta)=
\Sigma_n Z_0(ne^{-\tau})(a_n\cos n\theta
+b_n\sin n\theta),
\label{gensln}
\end{eqnarray}
where $Z_0$ is a solution to Bessel's equation
of order zero, and where the linearly growing
solution $P=P_0+\tau P_1$ ($P_0$ and $P_1$ constant)
has been suppressed.


A polarized Gowdy which evolves towards
a singularity is exemplified by
(Equation (4.2) in \cite{bv:a})
\begin{eqnarray}
P_0(\tau,\theta)=Y_0(e^{-\tau})\cos\theta,
\label{anal}
\end{eqnarray}
where $Y_0$ is the Bessel function of the second kind of
order zero.
With $Q=0$
the Hamiltonian constraints for $\lambda$
at $\tau=0$ become\cite{bv:a}
\begin{eqnarray}
\begin{array}{ll}
\lambda_\theta&=2P_\theta P_\tau,\\
\lambda_\tau  &=P^2_\tau+e^{-2\tau}P_\theta^2,
\end{array}
\label{gconB}
\end{eqnarray} 
so that (\ref{anal}) implies
\begin{eqnarray}
\begin{array}{ll}
\lambda(\tau,\theta)&=\frac{1}{2}Y_0(e^{-\tau})Y_1(e^{-\tau})e^{-\tau}
\cos2\theta+T(\tau),\\
T(\tau)&=\frac{1}{2}\int_{e^{-\tau}}^1
\{Y_0^{\prime 2}(t)+Y_0^2(t)\}t\mbox{d}t.
\end{array}
\label{lambda0}
\end{eqnarray}
Equations (\ref{anal}) and (\ref{lambda0})
provide for
initial data for simulations, 
an are an analytical reference solution 
for polarized waves.
The computational procedure 
to treat
the integral $T(\tau)$ 
is outlined in
the Appendix.


\subsection{Unpolarized Gowdy waves}

Unpolarized Gowdy waves follow from general
initial conditions on the $P(\tau,\theta)$
and $Q(\tau,\theta)$. Following the
earlier computations of Berger $et$ $el.$ \cite{bv:a},
we consider
\begin{eqnarray}
\begin{array}{ll}
P(0,\theta)=0, 
&\partial_\tau P(\tau,\theta)=A\cos(\theta),\\
Q(0,\theta)=
B \cos(\theta),
&\partial_\tau Q(\tau,\theta)=0,
\end{array}
\end{eqnarray}
where $A$ and $B$ are constants.
We then have the expansions
\begin{eqnarray}
\begin{array}{ll}
P(\tau,\theta)&\sim A\tau\cos\theta-\frac{1}{2}\tau^2B^2\sin^2\theta,\\
Q(\tau,\theta)&\sim (1-\frac{1}{2}B\tau^2)\cos\theta,\\
\lambda(\tau,\theta)&\sim \tau(A^2\cos^2\theta+B^2\sin^2\theta)
-\tau^2B^2\sin^2\theta
\end{array}
\end{eqnarray}
valid to second order in $\tau$. These expansions are used to 
compute initial data for the simulations.

\section{Integration scheme}

The integration scheme combines an evolution scheme for
the tetrad connections $\omega_{a\mu\nu}$ and
an evolution scheme for the tetrad legs
$(e_\mu)^b$.
The scheme is one-dimensional, second-order accurate in
the time-coordinate $\tau$ and spectrally accurate in the
space-coordinate $\theta$.
The scheme uses a straightforward leap frog time-stepping
algorithm, and,
in view of periodicity in $\theta$,
differentiation by the Fast Fourier Transform.
In this fashion, the errors are 
essentially due to the time-stepping algorithm.

In what follows, 
we shall work with both tensors and their
densities, denoted by
a tilde, $e.g.$,
$\tilde{\phi}=\sqrt{-g}\phi$. 
Time is discretized as $\tau_n=n\Delta \tau$,
indicated by a superscript: 
$\phi(\tau_n)=\phi^n$.
Indices $a,b$ will refer to all four space-time indices,
and $p,q,r,s,u,v$ to the 
spatial $\theta,\delta$ and $\sigma$ only.

\subsection{Evolution of the connections $\omega_{a\mu\nu}$}

The wave equations for $\omega_{a\mu\nu}$
are implemented directly through the divergence of the Riemann
tensor, its representation in terms of the connections and
the 
Lorentz 
gauge condition $c_{\mu\nu}=\nabla^a\omega_{a\mu\nu}=0$.

The integration is based on
the four-divergence of the Riemann tensor 
in the tetrad approach,
$\hat{\nabla}^aR_{ab\mu\nu}=0$. In coordinate notation, 
this gives 
$\partial_a\tilde{R}^{ab}_{\hskip.10in\mu\nu}
+[\omega_a,\tilde{R}^{ab}]_{\mu\nu}=0.$
Application of leap frog time-stepping gives the iterations
\begin{eqnarray}
\begin{array}{rl}
(\tilde{R}^{\tau r}_{\hskip.10in\mu\nu})^{n+1}
=&
(\tilde{R}^{\tau r}_{\hskip.10in\mu\nu})^{n-1}\\
&-2\Delta\tau(\partial_p
\tilde{R}^{p r}_{\hskip.10in\mu\nu}
+[\omega_a,\tilde{R}^{ar}]_{\mu\nu})^n.
\end{array}
\end{eqnarray}
Evolution equations for the connections 
$\omega_{p\mu\nu}$ (where $p$ is a spatial coordinate)
then follow from the identity
$R_{\tau p\mu\nu}
=\partial_\tau\omega_{p\mu\nu}-\partial_p\omega_{\tau\mu\nu}
+[\omega_\tau,\omega_p]_{\mu\nu}$. Application of leap frog
time-stepping gives
\begin{eqnarray}
\begin{array}{rl}
(\omega_{p\mu\nu})^{n+1}
=&
(\omega_{p\mu\nu})^{n-1}\\
&+2\Delta\tau(R_{\tau p\mu\nu}
+\partial_p\omega_{\tau\mu\nu}
-[\omega_\tau,\omega_p]_{\mu\nu})^n.
\end{array}
\end{eqnarray}
Evolution of the remaining $\omega_{\tau\mu\nu}$ are 
determined from the 
Lorentz gauge conditions
$c_{\mu\nu}=\nabla_c\omega^c_{\hskip.04in\mu\nu}=0,$ and hence
\begin{eqnarray}
\begin{array}{rl}
(g^{\tau\tau}\tilde{\omega}_{\tau\mu\nu})^{n+1}
&+(g^{\tau p}\tilde{\omega}_{p\mu\nu})^{n+1}
=(\tilde{\omega}^\tau_{\hskip.04in\mu\nu})^{n-1}\\
&-2\Delta \tau \partial_p(\tilde{\omega}^p_{\hskip.04in\mu\nu})^n.
\end{array}
\end{eqnarray}

Note that 
$R^{n}_{\tau p\mu\nu}$ 
appears in Step (b)
of the iteration scheme.
An update
$R^{n+1}_{\tau p\mu\nu}$ 
is algebraically related to 
$(R^{\tau p}_{\hskip0.07in\mu\nu})^{n+1}$
and $(R_{pq\mu\nu})^{n+1}$, the first of which follows from Step (a)
and the second of which follows from Step (b). 
Suppressing momentarily the tetrad idices $\mu\nu$ and the time-label
$n+1$, we have
\begin{eqnarray}
\begin{array}{lll}
R_{\tau p}&=&g_{\tau c}g_{dp}R^{cd}
=2g_{\tau [\tau }g_{r]p}R^{\tau r}+g_{\tau[s}g_{r]p}R^{sr}\\
&=&2g_{\tau [\tau }g_{r]p}R^{\tau r}+g_{\tau[s}g_{r]p}g^{sa}g^{br}R_{ab}\\
&=&
A_p^v R_{\tau v}+B_p,
\end{array}
\end{eqnarray}
were
\begin{eqnarray}
\begin{array}{rl}
A^v_p &=2g_{\tau [s}g_{r]p}g^{\tau s}g^{rv},\\
B_p   
&=2g_{\tau [\tau }g_{r]p}R^{\tau r}
+g_{\tau [s}g_{r]p}g^{us}g^{rv}R_{uv}
\end{array}
\end{eqnarray}
Note that $A_p^v$ and the second term in $B_p$ 
is due to a shift,
$g_{\tau p}$; in particular,
$A_p^v=0$ whenever 
$g_{\tau p}=0$.
It follows that 
\begin{eqnarray}
(\delta_p^v-A^v_p)R_{\tau v}
=B_p.
\label{id_R}\end{eqnarray}
As a system of 3x3 equations, this
is readily inverted.
With all quantities at 
$t_{n+1}=(n+1)\Delta \tau$,
(\ref{id_R})
determines 
updates $(R_{\tau p\mu\nu})^{n+1}$ 
from the results of the previous steps.

\subsection{Evolution of 
the tetrad legs $(e_\mu)^b$}

The tetrad legs satisfy the equations of
structure
\begin{eqnarray}
\partial_{[a}(e\mu)_{b]}=(e_\nu)_{[b}\omega_{a]\mu}^{\hskip.17in\nu}.
\end{eqnarray}
Clearly, these equations leave 
the tetrad lapse functions $N_\mu=(e_\mu)_\tau$
as free variables. They are related to 
the Hamiltonian lapse and shift functions
$(N,N_p)$ through
\begin{eqnarray}
g_{a\tau}=N_\alpha(e^\alpha)_a=(N_pN^p-N^2,N_p).
\end{eqnarray}
These tetrad lapse functions
govern the evolution of the tetrad legs
in the equations of structure:
\begin{eqnarray}
\partial_\tau(e_\mu)_b+\omega_{\tau\mu}^{\hskip.1in\nu}(e_\nu)_b
=\partial_bN_\mu+\omega_{b\mu}^{\hskip.1in\nu}N_{\nu}
\equiv\hat{\partial}_bN_{\mu}.
\label{struc}
\end{eqnarray}
Thus, conditions on $g_{a\tau}$ from the line-element
(\ref{gmetr}) result in a system
of implicit equations
for the lapse functions $N_\mu$, as discussed in Section IV.

Integration of (\ref{struc}) can be written using
the fundamental matrix $\Lambda_\mu^{\hskip.04in\nu}(\tau;\tau_0)$:
a finite Lorentz transformation satisfying 
\begin{eqnarray}
\left\{
\begin{array}{ll}
\Lambda_\mu^{\hskip.04in\nu}(\tau,\tau)&=\delta_\mu^\nu,\\
\partial_\tau
\Lambda_\mu^{\hskip.04in\nu}(\tau;\tau_0)
&=-\omega_{\tau\mu}^{\hskip.1in\alpha}
(\tau)
\Lambda_\alpha^{\hskip.04in\nu}(\tau;\tau_0),
\end{array}
\right.
\label{def_l}
\end{eqnarray}
with which
the solution to (\ref{struc}) is
\begin{eqnarray}
\begin{array}{rl}
(e_\mu)_b(\tau)
=&
\Lambda_\mu^{\hskip.04in\nu}(\tau;\tau_0)(e_\nu)_b(\tau_0)\\
&+\int_{\tau_0}^{\tau}
\Lambda_{\mu}^{\hskip.04in\nu}(\tau;s)
\hat{\partial}_bN_\nu(s)
\mbox{d}s.
\label{formal}
\end{array}
\end{eqnarray}
Here, only the $\tau-$dependence has been made explicit.
Indeed, if the convolution
integral on the right hand-side were to vanish, the evolution of the tetrad
becomes a pure SO(3,1,$\bf R$)-gauge transformation, 
leaving the metric
$g_{ab}(\tau)=(e_\alpha)_a(\tau)(e^\alpha)_b(\tau)$ constant.
The general representation (\ref{formal}) shows
explicitly that the evolution of the 
metric
takes place in the (flat) tangent bundle of the four-dimensional,
physical manifold. The integral in (\ref{formal}) 
is implemented using Gaussian integration, including
a Taylor expansion
for the integrand as provided by the updates in Section 3.1.
Thus, in the case of given connections, the tetrad evolution
is carried out with machine precision.
In general, a Taylor expansion of $\omega_{a\mu\nu}$ up to its
$n-$th derivatives at $t_n$ in (\ref{formal}) provides
an update of the tetrad legs with $n+1-$th order accuracy.

\subsection{Equations for $\Lambda_\mu^{\hskip.04in\nu}(\tau;\tau_0)$}

In the Gowdy line-element, $\Gamma_{\tau c}^d$,
$\Gamma_{\theta c}^d$ and 
$R_{\tau\theta c}^{\hskip.15in d}$ all have a 2$\times$2 
block diagonal structure as matrices in $c,d$.
Initial data for the tetrad (Section D) can be chosen
with $(e_\mu)_b$ likewise, when regarded
as a matrix in $\mu,b$. This obtains
$\omega_{\tau\mu\nu}$,
$\omega_{\theta\mu\nu}$, $[\omega_\tau,\omega_\theta]_{\mu\nu}$
and $R_{\tau\theta\mu\nu}$ in block diagonal form at
$\tau=0$.
Consider now the evolution problem for the connections
in terms of $\partial_\tau\omega_{\theta\mu\nu}=
\partial_\theta\omega_{\tau\mu\nu}+R_{\tau\theta\mu\nu}
-[\omega_\tau,\omega_\theta]$
and the Lorentz gauge, $c_{\mu\nu}=0$.
Here,
$R_{\tau\theta\mu\nu}=R_{\tau\theta c d}(e_\mu)^c(e_\nu)^d$,
where (the structure of) $R_{\tau\theta cd}$ is
as mentioned above,
and where the tetrad elements satisfy (\ref{struc}) with
$N_\mu=(*,*,0,0)$ (Section IV). 
Consistent evolution is obtained in which the block diagonal
structure of the forementioned tensor elements is preserved
with
\begin{eqnarray}
\omega_{\tau\mu}^{\hskip.1in\nu}=
\left(
\begin{array}{lll}
a\sigma & 0\\
0 & b\epsilon
\end{array}
\right),
\label{def_a}
\end{eqnarray}
where $a=a(\tau,\theta)$ and $b=b(\tau,\theta)$, and
\begin{eqnarray}
\sigma=
\left(
\begin{array}{lll}
0 & 1\\
1& 0
\end{array}
\right),
\epsilon=
\left(
\begin{array}{lll}
0 & -1\\
1& 0
\end{array}
\right).
\end{eqnarray}
By explicit exponentiation, the Lorentz transformation 
of boosts on 
the $(e_T)^b$ and $(e_\Theta)^b$,
and rotations on
$(e_\Sigma)^b$ and 
$(e_\Delta)^b$
is therefore
\begin{eqnarray}
\Lambda_\mu^{\hskip.04in\nu}=
\left(
\begin{array}{ll}
K & 0\\
0 & L
\end{array}
\right).
\end{eqnarray}
Here, we adapt the nomenclature of \cite{jackson:a},
with the 2x2 matrices $K$ and $L$ given by
\begin{eqnarray}
\begin{array}{rl}
K_\mu^{\hskip.04in\nu}&=
\left(
\begin{array}{ll}
\cosh\phi &\sinh\phi\\
\sinh\phi &\cosh\phi
\end{array}
\right)\mbox{  }
(\mu,\nu=T,\Theta),\\
L_\mu^{\hskip.04in\nu}&=
\left(
\begin{array}{lr}
\cos\psi &-\sin\psi\\
\sin\psi &\cos\psi
\end{array}
\right)\mbox{  }
(\mu,\nu=\Sigma,\Delta)
\end{array}
\end{eqnarray}
with
\begin{eqnarray}
\phi(\tau;\tau_0)=\int^{\tau_0}_\tau a(s)\mbox{d}s,
\mbox{  }\psi(\tau;\tau_0)=\int^{\tau_0}_\tau b(s)\mbox{d}s.
\label{PHI}
\end{eqnarray}

\subsection{Initial data and choice of tetrad}

The integration scheme requires 
$\omega_{a\mu\nu}$ and
$\partial_\tau\omega_{a\mu\nu}$ for obtaining initial values of
the Riemann tensor on 
$\tau=0$.
For the polarized Gowdy wave 
($Q=0$), we can simply use 
\begin{eqnarray}
\begin{array}{lll}
(e_T)^b&=(e^{\frac{\lambda}{4}},0,0,0),\\
(e_\Theta)^b&=(0,e^{\frac{\lambda}{4}},0,0),\\
(e_\Sigma)^b&=(0,0,e^{-\frac{P_0}{2}},0),\\
(e_\Delta)^b&=(0,0,0,e^{\frac{P_0}{2}}).
\end{array}
\label{tetr0}
\end{eqnarray}

In case of the full Gowdy metric ($P,Q\ne0$),
care must be given to the choice
of initial tetrad 
to ensure 
sufficient
differentiability for
obtaining smooth 
connections $\omega_{a\mu\nu}$ and 
Riemann tensor $R_{ab\mu\nu}$.
The tetrad legs for the $\tau-\theta$
coordinate plane can be chosen 
as before,
\begin{eqnarray}
(e_T)^b&=(e^{\frac{\lambda}{4}},0,0,0),\\
(e_\Theta)^b&=(0,e^{\frac{\lambda}{4}},0,0).
\end{eqnarray}
The tetrad legs for the $\sigma-\delta$
coordinate plane,
\begin{eqnarray}
E=\left(
\begin{array}{lr}
(e_\Sigma)_\sigma&(e_\Delta)_\sigma\\
(e_\Sigma)_\delta&(e_\Delta)_\delta
\end{array}
\right),
\end{eqnarray}
must satisfy $EE^T=G$, $EG^{-1}E^T=I$,
where $I$ is the 2x2 identity matrix and
$G$ is the matrix with the metric components
\begin{eqnarray}
G=\left(
\begin{array}{ll}
g_{\sigma\sigma}&g_{\sigma\delta}\\
g_{\sigma\delta}&g_{\delta\delta}
\end{array}
\right)=
e^{-\tau+P}\left(
\begin{array}{ll}
1&Q\\
Q&Q^2+e^{-2P}
\end{array}
\right).
\end{eqnarray}
$E$ can be defined uniquely as the symmetric positive
definite square root ($cf.$ \cite{strang:a}) of $G$.
To this end, let
\begin{eqnarray}
A(\phi)=\left(
\begin{array}{lr}
\cos\phi & -\sin\phi\\
\sin\phi & \cos\phi
\end{array}
\right)
\end{eqnarray}
denote both the matrix containing
the tangents to the tetrad legs, and
the rotation matrix on the tetrad indices
$\mu=\sigma,\delta$. Then
\begin{eqnarray}
E=A(\phi)\left(
\begin{array}{lr}
\lambda_+^{\frac{1}{2}}&0\\
0&\lambda_-^{\frac{1}{2}}
\end{array}
\right)
A(-\phi),
\label{tE}
\end{eqnarray}
where the eigenvalues
$\lambda_\pm$ of $G$ can be given in terms
of $z=\frac{1}{2}(Q^2+e^{-2P}-1)$
and $r=\sqrt{Q^2+z^2}$ as
\begin{eqnarray}
\lambda_\pm=(1+z\pm r)e^{-\tau+P}.
\end{eqnarray}
The rotation angle $\phi$ follows from the
equations
\begin{eqnarray}
z\sin2\phi+Q\cos2\phi=0,
\mbox{   }z+r\cos2\phi=0.
\label{EqnPhi}
\end{eqnarray}
The second equation in (\ref{EqnPhi})
gives $\cos2\phi=-\frac{z}{r}$, and
hence $\sin2\phi=\frac{Q}{r}$.
Without loss of generality,
$\cos\phi\ge0$, so that
$0\le\phi\le\frac{\pi}{2}$
if $Q\ge0$, and 
$-\frac{\pi}{2}\le\phi<0$ if $Q<0$.
The additional rotation 
$A(-\phi)$ on the tetrad indices 
in the right hand-side of (\ref{tE})
effectively regularizes
$E$ as $G$
approaches the 2x2 identity
matrix (when $P$ and $Q$
become small, in which case
$\phi$ becomes ill-defined).

The connections $\omega_{a\mu\nu}$ are now computed using
its metric definition
\begin{eqnarray}
\omega_{a\mu\nu}=(e_\mu)_c\partial_a(e_\nu)^c
+\Gamma_{ab}^{c}(e_\mu)_c(e_\nu)^b.
\label{ndefom}
\end{eqnarray}
Initial data
for the Riemann tensor 
of Section IIIa
now follow.


\section{Gauge conditions for Gowdy wave}

The Gowdy line-element prescribes a certain
slicing of space-time, which are incorporated in the
simulations to enable comparison with
the analytical solution for the polarized and
pseudo-unpolarized Gowdy waves, and the
reference solutions obtained by integration of the
reduced wave equations (\ref{gwave}) and
Hamiltonian constraints (\ref{gcon}).

The Gowdy line-element (\ref{gmetr}) has algebraic slicing conditions
\begin{eqnarray}
\left\{
\begin{array}{ll}
g_{\tau\tau}&=-e^{-2\tau}g_{\theta\theta},\\
g_{\tau\theta}&=0,\\
g_{\tau\sigma}&=0,\\
g_{\tau\delta}&=0.
\end{array}
\right.
\label{gauge_gowdy}
\end{eqnarray}
We have
\begin{eqnarray}
\begin{array}{lllll}
g_{\tau\tau}&=&
(e_\alpha)_\tau(e^\alpha)_\tau&=&-N_T^2+N_\Theta^2,\\
g_{\theta\theta}&=&
(e_\alpha)_\theta(e^\alpha)_\theta&=&
-(e_T)^2_\theta+(e_\Theta)^2_\theta,\\
g_{\tau\theta}&=&(e_\alpha)_\tau(e^\alpha)_\theta&=&
-N_T(e_T)_\theta+N_\Theta(e_\Theta)_\theta.
\end{array}
\end{eqnarray}
The shift condition $g_{\tau\theta}=0$ gives
\begin{eqnarray}
\frac{N_\Theta}{N_T}=
\frac{(e_T)_\theta}{(e_\Theta)_\theta},
\end{eqnarray}
so that
\begin{eqnarray}
g_{\theta\theta}=
(e_\Theta)^2_\theta
\left(1-\frac{(e_T)^2_\theta}{(e_\Theta)^2_\theta}\right)
=(e_\Theta)^2_\theta
\left(1-\frac{N_\Theta^2}{N_T^2}\right).
\end{eqnarray}
Because also
\begin{eqnarray}
g_{\tau\tau}=
-N_T^2\left(1-\frac{N_\Theta^2}{N_T^2}\right),
\end{eqnarray}
the lapse condition $g_{\tau\tau}=-e^{-2\tau}g_{\theta\theta}$
becomes
\begin{eqnarray}
N_T^2=e^{-2\tau}(e_\Theta)_\theta^2.
\end{eqnarray}
We thus have the following conditions on the tetrad lapse functions
\begin{eqnarray}
\left\{
\begin{array}{ll}
N_T&=-e^{-\tau}(e_\Theta)_\theta
=-e^{-\tau}\sigma_T^{\hskip0.04in\gamma}
(e_\gamma)_\theta,\\
N_\Theta&=-e^{-\tau}(e_T)_\theta
=-e^{-\tau}\sigma_\Theta^{\hskip0.04in\gamma}
(e_\gamma)_\theta,\\
N_\Sigma&=0,\\
N_\Delta&=0.
\end{array}
\right.
\label{lapses}
\end{eqnarray}
Upon substitution of (\ref{lapses}) into the
equations of structure (\ref{formal}), 
two implicit equations for the $(e_\Theta)_\theta$
and $(e_T)_\theta$ result.

\subsection{Solution of gauge conditions}

The nontrivial $\tau-\theta$ gauge conditions in (\ref{lapses})
are separated from 
the trivial gauge conditions on
the $\sigma-\delta$ coordinates.
Our starting point, therefore, is 
\begin{eqnarray}
(e_\mu)_\theta(\tau,\theta)
=K_\mu^{\hskip.04in\nu}(\tau;\tau_0,\theta)
\eta_{\nu}(\tau,\theta),
\mbox{  ($\mu,\nu=T,\Theta)$},
\label{rep_e}
\end{eqnarray}
and to work in the $\tau-\theta$ sector only.
In what follows, Greek tetrad indices
run through $T,\Theta$.
Also, the spatial ($\theta-$)dependence will made explicit only
when needed.

The original gauge conditions (\ref{gauge_gowdy})
are given in terms of the metric,
which are invariants under pointwise 
$K$ transformations applied 
simultaneously to both $(e_T)^b$
and $(e_\Theta)^b$.
Upon substitution of (\ref{rep_e}) in
(\ref{lapses}), therefore, $K$ 
can be factored out, 
leaving a linear 
equation for $\eta_\mu$.

Expressed in (\ref{rep_e}), the gauge conditions (\ref{lapses})
become
\begin{eqnarray}
N_\mu=-e^{-\tau}\sigma_\mu^{\hskip.04in\nu}K_\nu^{\hskip.04in\gamma}\eta_\gamma.
\label{lapses2}
\end{eqnarray}
Substitution of (\ref{rep_e}) and (\ref{lapses2}) into the 
equations of
structure (\ref{struc})
gives two implicit equations for $\eta_\mu=(\eta_T,\eta_\Theta)$.
Note that
\begin{eqnarray}
\begin{array}{ll}
\omega_{\theta\mu}^{\hskip.10in\nu}(\tau,\theta)&
=\beta(\tau,\theta)\sigma_\mu^{\hskip.04in\nu}\\
\partial_\theta K_\mu^{\hskip.04in\nu}(\tau;\tau_0,\theta)
&=\sigma_\mu^{\hskip.04in\gamma}K_\gamma^{\hskip.04in\nu}(\tau;\tau_0,\theta)
\alpha(\tau;\tau_0,\theta),
\end{array}
\label{def_ba}
\end{eqnarray}
where $\beta=\beta(\tau,\theta)$ 
is a coefficient function, and
$\alpha(\tau;\tau_0,p)=\int_{\tau}^{\tau_0}\partial_\theta
a(s,p)\mbox{d}s$.
Together with 
$\partial_tK_\mu^{\hskip.04in\nu}
=-a\sigma_{\mu}^{\hskip.04in\alpha}K_\alpha^{\hskip.04in\nu}$
from the second equation
in (\ref{def_l}), and the algebraic properties
$\sigma_\mu^{\hskip.04in\gamma}\Lambda_\gamma^{\hskip.04in\nu}
=\Lambda_\mu^{\hskip.04in\gamma}
\sigma_\gamma^{\hskip.04in\nu}$
and
$\sigma_\mu^{\hskip.04in\gamma}
\sigma_\gamma^{\hskip.04in\nu}=\delta_\mu^\nu$,
the linear equation
for $\eta_\mu$ follows:
\begin{eqnarray}
\partial_\tau\eta_\mu
+e^{-\tau}
(\sigma_\mu^{\hskip.04in\nu}\partial_\theta\eta_{\nu}
 +c(\tau,\tau_0)\eta_\mu)=0,
\label{dif_eta}
\end{eqnarray}
where 
$c(\tau,\tau_0):=\beta(\tau)+\alpha(\tau;\tau_0)$. 
Because $\alpha(\tau;\tau)\equiv0$,
$c(\tau,\tau_0)$ combines
$\omega_{\theta\mu}^{\hskip.1in\nu}$ 
with $\partial_\theta\omega_{\tau\mu}^{\hskip.1in\nu}$ to first and second order in
$(\tau-\tau_0)$, respectively.

Momentarily suppressing the tetrad index $\mu$,
we now look for a solution to (\ref{dif_eta}) in
\begin{eqnarray}
\eta=\eta_0+(\tau-\tau_0)\eta_1+\frac{1}{2}(\tau-\tau_0)^2\eta_2
+\frac{1}{3}(\tau-\tau_0)^3\eta_3+\cdots,
\label{exp_eta}
\end{eqnarray}
where each $\eta_k=\eta_k(\theta)$ has only 
$\theta-$dependence.
Similarly, we write
\begin{eqnarray}
\begin{array}{rl}
e^{\tau}&=e^{\tau_0}+
(\tau-\tau_0)
e^{\tau_0}
+\frac{1}{2!}
(\tau-\tau_0)^2
e^{\tau_0}
+\cdots,\\
c(\tau,\tau_0)&=c_0+(\tau-\tau_0)c_1(\tau_0)
+(\tau-\tau_0)^2c_2(\tau_0)+\cdots.
\end{array}
\label{exp_ec}
\end{eqnarray}
Expansions (\ref{exp_eta}) and (\ref{exp_ec})
can substituted into (\ref{dif_eta}); 
matching coefficients gives
\begin{eqnarray}
\sigma_\mu^{\hskip.04in\nu}\partial_\theta
\eta_{\nu 0}+c_0\eta_0=0
\end{eqnarray}
for $k=0$, and
\begin{eqnarray}
e^{\tau_0}\sum_{l=0}^{k}\frac{1}{l!}\eta_{k+1-l}
+\frac{1}{k}\sigma_\mu^{\hskip.04in\nu}\partial_\theta
\eta_{\nu k}+c_k\eta_0
+\sum_{l=0}^{k-1}\frac{1}{k-l}c_l\eta_{k-l}=0,
\end{eqnarray}
for $k\ge1$, $i.e.$,
\begin{eqnarray}
\begin{array}{rl}
\eta_{k+1}+
&\sum_{l=1}^{k}\frac{1}{l!}\eta_{k+1-l}
+e^{-\tau_0}
(\frac{1}{k}\sigma_\mu^{\hskip.04in\nu}\partial_\theta
\eta_{\nu k}\\
&+c_k\eta_0+
\sum_{l=0}^{k-1}\frac{1}{k-l}c_l\eta_{k-l})=0.
\end{array}
\end{eqnarray}
The first few terms are
\begin{eqnarray}
\left\{
\begin{array}{rl}
\eta_{\mu0}=&
(e_\mu)_\theta(\tau_0),\\
\eta_{\mu1}=&
-e^{-\tau_0}
(\sigma_\mu^{\hskip.04in\nu}\partial_\theta\eta_{\nu0}
+c_0\eta_{\mu0}),\\
\eta_{\mu2}=&-\eta_{\mu1}
-e^{-\tau_0}
(\sigma_\mu^{\hskip.04in\nu}\partial_\theta\eta_{\nu1}
+c_1\eta_{\mu0}+c_0\eta_{\mu1}),\\
\eta_{\mu3}=&-\eta_{\mu2}-\frac{1}{2!}\eta_{\mu1}\\
            &-e^{-\tau_0}
(\frac{1}{2}\sigma_\mu^{\hskip.04in\nu}\partial_\theta\eta_{\nu2}
+c_2\eta_{\mu0}+c_1\eta_{\mu1}
+\frac{1}{2}c_0\eta_{\mu2}),\\
\cdots=&\cdots.
\end{array}
\right.
\end{eqnarray}
This series expansion has been implemented numerically
using a numerical cut-off of 1.D-12, thereby maintaining
the Gowdy line-element within machine accuracy.

\section{Results and conclusions}

The performance of the numerical implementation
has been studied by varying both space and time
discretization. 
Verification of second-order accuracy has been
obtained by computing the ratio of the errors
(as a function of time) of those with $2m$
time-steps to those with $m$ time-steps for
a given final value of $\tau$. In all computations, a moderate
degree of discretization in the $\theta-$variable has been
found adequate, because of the spectral accuracy
in $\theta-$differentiation by FFT.
The computations have been performed in the Gowdy line-element,
using the freedom of space-time slicing in the present
formulation.
Numerical results and convergence data have been
obtained for the polarized, pseudo-unpolarized and
the two unpolarized Gowdy waves. 
The results are shown in Figs. 1-3, for both
the polarized wave (Figure 1) and two unpolarized waves
(Figure 2,3).
The results 
show second-order convergence in 
accord with the numerical scheme.
Also shown is convergence in the errors in the Ricci tensor
(also second-order). 
The results suggest that more advanced time-stepping
algorithms should be applicable, to adapt for
for more accurate, long-time computations.
At this stage, the results do not indicate a need for implicit
time-stepping. 

The computational problem of binary coalescence of neutron
stars or black holes requires a continuing development,
including an extension to three-dimensions,
adaptation to non-periodic
boundary conditions, and possibly a minimization of the error
in the Ricci tensor. 

Two directions for further exploration
stand out, which are motivated by the structure of the equations.
Firstly, the present formulation offers the first
strictly-hyperbolic formulation with complete freedom
of foliation (in particular, there is no restriction on the 
equivalent Hamiltonian lapse function). It will be of
interest to exploit this for new avenues in the treatment of 
horizon boundary conditions, and possibly in the extraction
of wave forms at the outer boundary as well.
Secondly, the implementation is based on the four-divergence 
of the Riemann tensor and contains no Christoffel symbols.
Somewhat analogous equations are given by Faraday's
equations, which recently have been successfully
implemented using cylindrical coordinates in
simulations of magnetized relativistic jets
\cite{mvp:94a,mvp:apj96}. 
Here, the notorious axis instabilities are fully regularized
\cite{mvp:BC}, 
whence it will be of interest to extend
this regularization to the present equations.
Cylindrical coordinates would be desirable 
in view of the expected simplification in the asymptotic wave 
form at large distances.

Finally, inclusion of a non-trivial stress-energy tensor is 
required in the problem of coalescence of neutron stars
and possibly black holes as well \cite{mvp:96b}. 

\mbox{}\\
\mbox{}\\
{\bf Acknowledgment.}
The author greatfully acknowledges stimulating discussions with
Saul Teukolsky.
This work has been supported in part by NSF grant 94-08378 and
by the Grand Challenge grant  NSF PHY 93-18152/ASC 93-18152
(ARPA supplemented). The Cornell Theory Center is supported by
NSF, NY State, ARPA, NIH, IBM and others.

\mbox{}\\
\mbox{}\\
{\bf Appendix.}
The analytical solution
of the polarized Gowdy wave 
is given by $P_0(\tau,\theta)$
in (\ref{anal}) and $\lambda(\tau,\theta)$ in (\ref{lambda0}).
It has been found useful in debugging the program and
in studying convergence to be able to monitor each and
every variable during the simulation in comparison with the
exact solution. To obtain these and all derived quantities, such as 
$\omega_{a\mu\nu}$ and $R_{ab\mu\nu}$, at every time-step,
numerical evaluation of the right hand-sides of (\ref{anal})
and (\ref{lambda0}) are needed.
A convenient method to do so is by a direct evaluation of their
defining expressions in terms of the metric, using numerical
differentiation by finite differences across 
small time- and space-intervals
$\epsilon$ (much smaller than the 
time-step size and spatial discretization in the numerical
integration of the wave equations).
This requires accurate evaluation of
\begin{eqnarray}
\begin{array}{rl}
T(\tau)&=\frac{1}{2}\int^1_{e^{-\tau}}
\{Y^{\prime 2}_0(t)+Y^2_0(t)\}t\mbox{d}t\\
       &=\frac{1}{2}\int_0^\tau
\{Y^{\prime 2}_0(e^{-\tau})+Y^2_0(e^{-\tau})\}e^{-2\tau}\mbox{d}\tau
\end{array}
\end{eqnarray}
up to its second derivatives about each of the 
$\tau_n=n\Delta\tau$.

Generally, $T(\tau)$ on $[\tau_a,\tau_b]$
is given accurately up to its $k$-th derivatives in each
subinterval 
$D_n=[\tau_n-k\epsilon,\tau_n+k\epsilon]$
about $\tau_n=n\Delta\tau=\frac{\tau_b-\tau_a}{m}$ by
a chain of function elements $\{f_n,D_n\}$ is used 
(see, $e.g.$ \cite{ru:a}).
Here, the $f_n$ are polynomial approximations 
\begin{eqnarray}
f_n(\tau)=c_{n0}+c_{n1}(\tau-\tau_n)+c_{n2}(\tau-\tau_n)^2+
\cdots
\end{eqnarray}
to $T(\tau)$ on $D_n$, and  
the $D_n$ cover $[-k\epsilon,m\Delta\tau+k\epsilon]$.
Naturally, the degree of the $f_n$ is chosen to be 
greater than or equal to $k$.
In the case at hand, $k=2$,
and the degree of $f_n$ is chosen to be three.

The coefficients $c_{nl}$, $l=1,2,\cdots$ follow by direct evaluation of
the (analytic) prescription
of the integrand of $T(\tau)$ and its (analytical) derivatives. 
A set of accurate values of the $c_{n0}$ then follows by first evaluating 
$\{T(\tau^\prime_k)\}$ on
\begin{eqnarray}
\tau_k^\prime=\frac{1}{2}m\Delta\tau^\prime
(1+\cos x_k), \mbox{  }x_k=\frac{2\pi}{N},
\end{eqnarray}
using integration of its integrand
on the $\{\tau_k^\prime\}$ by the Fast Fourier Transform,
where $N$ is the
degree of the FFT, followed by an 
interpolation of the $T(\tau_k^\prime)$
to the $T(\tau_n)$ using the $c_{nl}$, $l=1,2,\cdots$.

In our computations $N=256$, which 
gives results on $T(\tau)$ to within machine accuracy.
Of course, for every derived quantity involving numerical
differentiation, there is a loss of accuracy 
determined by the numerical differentiation parameter $\epsilon$.
We have chosen $\epsilon=10^{-6}$.


\begin{thebibliography}{10}

\bibitem{bradaschia:a}
C.~Bradaschia{,} E. Caloni{,} M. Cobal{,} R. Del Fasbro{,} A. Di Virgilio{,} A.
  Giazotta{,} E. Holloway{,} H. Kautzky{,} B. Michelozzi{,} V. Montelatici{,}
  D. Pascuello{,} W.~Velloso{,} in~{\em Gravitation 1990}{,} Proceedings of the
  Banff Summer Institute{,} Banff{,} Alberta{,} edited~by R.~Mann \& P. Wesson
  (World Scientific{,} Singapore{,}~1991).

\bibitem{jackson:a}
Jackson.
\newblock {\em CLassical ELectrodynamics}.
\newblock John Wiley \& Sons, New York, 1975.

\bibitem{bv:a}
B.K. Berger \&~V. Moncrief.
\newblock {\em Phys. Rev. D}, 48(10):4676--4687, 1993.

\bibitem{strang:a}
G.~Strang.
\newblock {\em Linear Algebra and its applications}.
\newblock Academic Press, New York, 1980.

\bibitem{mvp:94a}
M.H.P.M. van Putten.
\newblock Proc. Cornelius Lanczos Int. Centenary Conf., edited by J.D. Brown,
  M.T. Chu, D.C. Ellison and R.J. Plemmons, p449 (SIAM, Philadelphi, 1994).

\bibitem{mvp:BC}
M.H.P.M. van Putten.
\newblock {\em Int. J. Bifurcation \& Chaos}, 4(1):57--69, 1994.

\bibitem{mvp:apj96}
M.H.P.M. van Putten.
\newblock 467:L57, 1996.

\bibitem{mvp:96b}
van Putten~M.H.P.M.
\newblock {\em Phys. Rev. D}, 54(10):R5931, 1996.

\bibitem{mvp:IX}
van Putten M.H.P.M. \& Douglas M.~Eardley.
\newblock {\em Phys. Rev. D}, 53(6):3056, 1996.

\bibitem{ru:a}
Ruding W.
\newblock {\em Real \& Complex Analysis}.
\newblock McGraw-Hill, New York, 1987.

\bibitem{abramovici:a}
A.~Abramovici{,} W.E. Althouse{,} R.W.P. Drever{,} Y.Gursel{,} S.Kawamura{,}
  F.J. Raab{,} D. Shoemakes{,} L. Siewers{,} R.E. Spero{,} K.S. Thorne{,} R.E.
  Vogt{,} R. Weis{,} S.E. Whitcomb \&~M.E. Zucker.
\newblock {\em Science}, 256:325, 1992.

\end{thebibliography}

\newpage
\begin{figure}
\caption
{Shown is the simulation on $0\le\tau\le5.12$
of the polarized
Gowdy wave.
Distributions $\lambda(\tau,\theta)$ and 
$P(\tau,\theta)$ 
are displayed (upper windows), together with their
$\theta-$distribution at $\tau=5.12$
(middle windows).
Errors are obtained in computations with 
a consecutive doubling 
of the 
number of time-steps ($m=512,1024,2048$), and
are given as a function of
time in the lower window to the left for both
$P$ and $\lambda$ 
(o for $m=512$, $\times$ for $m=1024$ and
lines for $m=2048$). The errors show proper
second order convergence 
(lower window to the right), obtained
upon dividing the errors for 
$m=1024$ by those of $m=512$
($\times$), and
those for $m=2048$ by those of $m=1024$ (lines).
These errors have been computed with reference to
the analytical solution to 
Gowdy's reduced wave equations.} 
\end{figure}
 
\begin{figure}
\caption
{Shown is the 
simulation on
$0\le\tau\le5.12$ of the unpolarized Gowdy wave in
response to initial data with $A=B=1$.
Discretization in $\theta$ is $n=64$
(32 points displayed).
Errors are obtained in computations with 
a consecutive doubling 
of the 
number of time-steps ($m=512,1024,2048$), and
are given as a function of
time in the lower window to the left for each 
of the three functions
$P$, $Q$ and $\lambda$ 
(o for $m=512$, $\times$ for $m=1024$ and
lines for $m=2048$). The errors show proper
second order convergence 
(lower window in the middle), obtained
upon dividing the errors for 
$m=1024$ by those of $m=512$
($\times$), and
those for $m=2048$ by those of $m=1024$ (lines).
The errors in the Ricci tensor for the
three discretizations (lower window to the right) likewise 
show second order convergence. An additional dashed line
further shows the error in the Ricci tensor for extremely high 
time-discretization,
$m=32768$. These errors are computed using 
numerical results from 
Gowdy's reduced wave 
equations for comparison.}
\end{figure}

\begin{figure}
\caption
{Shown is the simulation on $0\le\tau\le7.68$ of the unpolarized Gowdy wave in
response to initial data with $A=0$ and $B=1$.
Discretization in $\theta$ is $n=64$ (32 points displayed).
The windows display the same 
variables and errors as described
in the previous figure, with errors again computed using
results from Gowdy's 
reduced wave equations for 
comparison. A notable transition occurs in the behavior of the
errors, specifically in those of the Ricci tensor, as steep gradients
are formed in
$P$ and $Q$ (and to a lesser
extend in $\lambda$ as well)
about $\tau=5$.}
\end{figure}
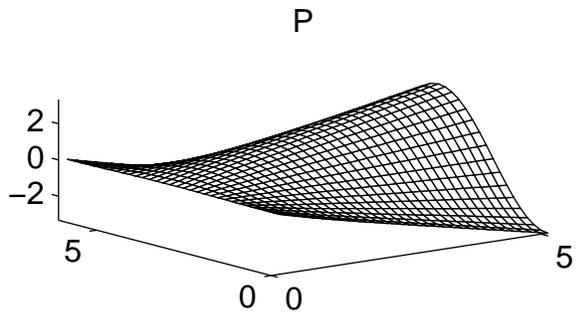
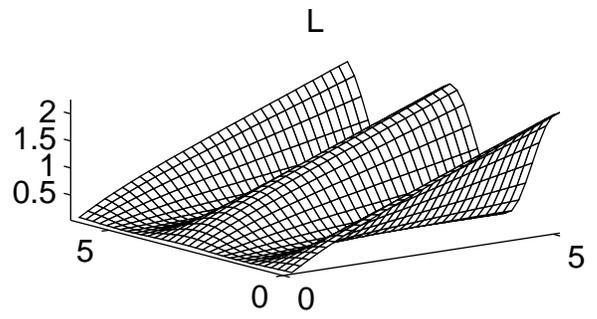
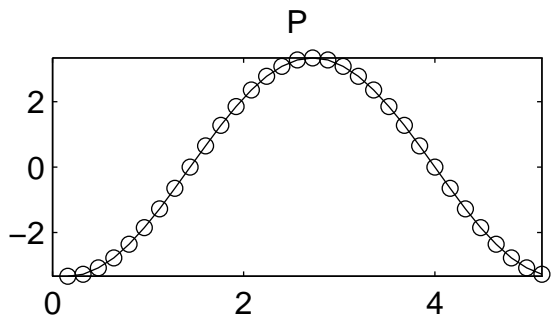
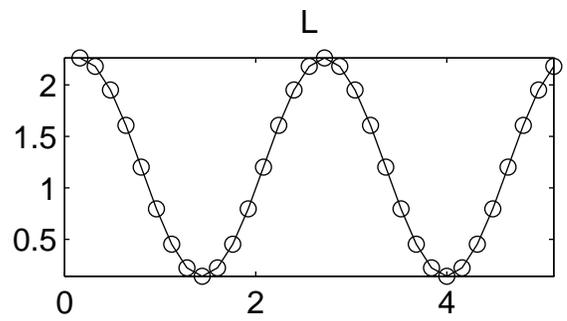
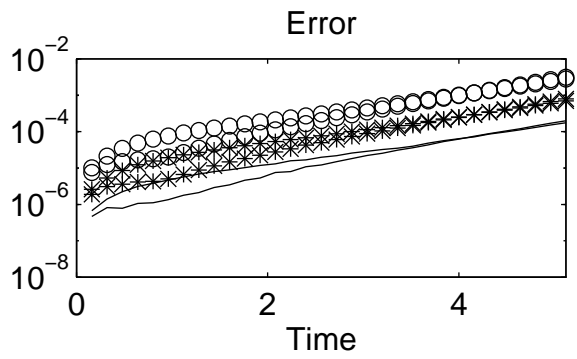
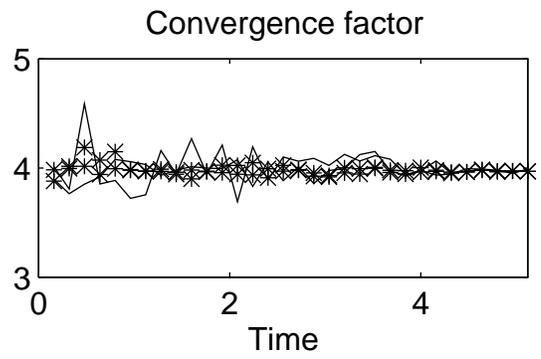
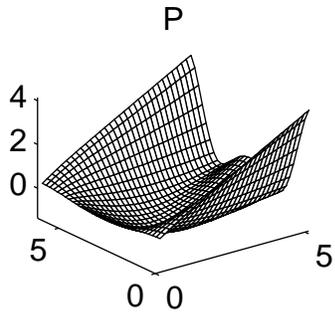
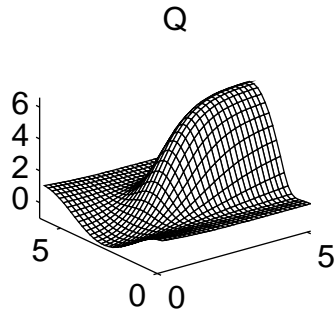
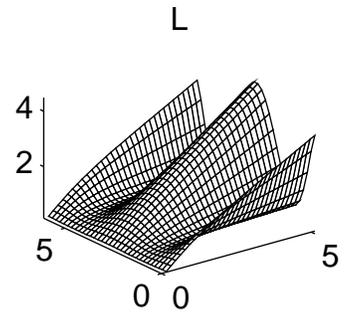
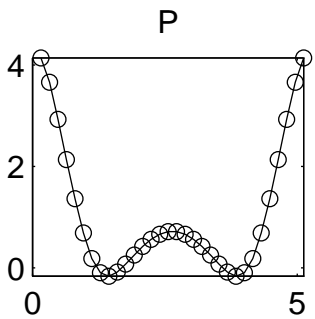
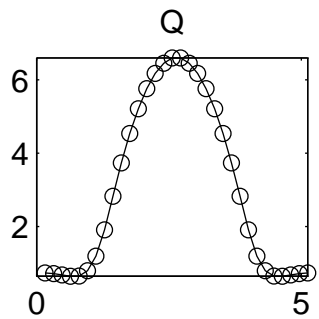
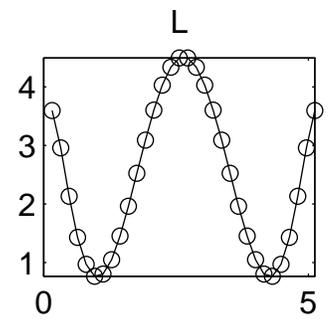
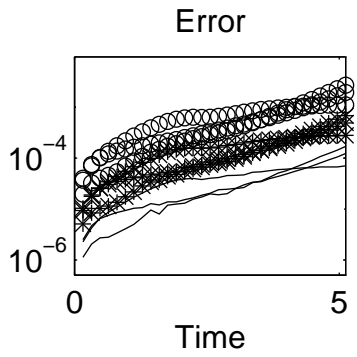
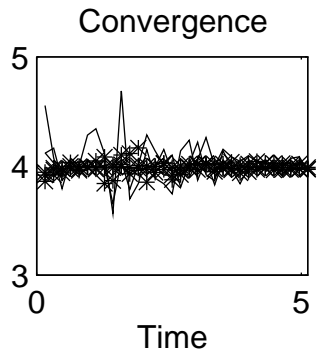
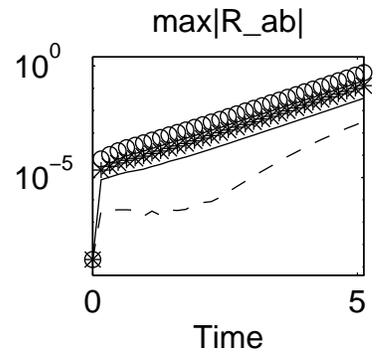
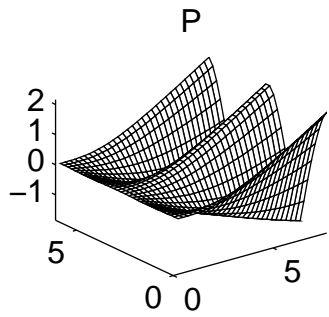
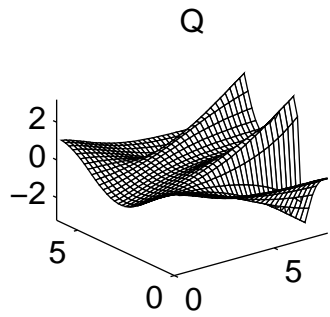
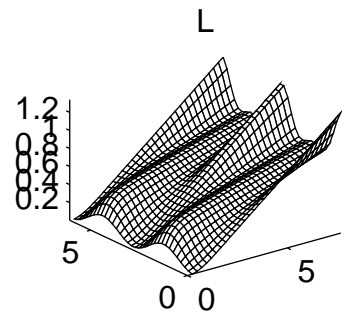
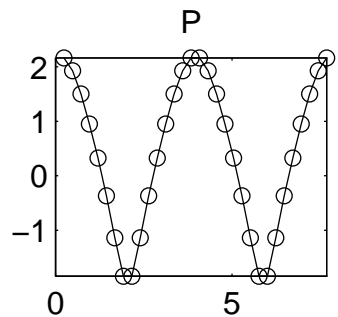
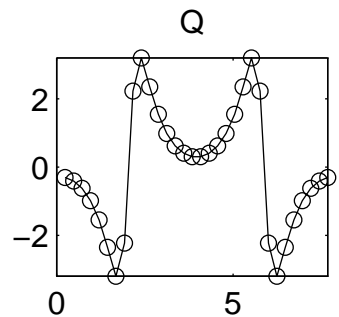
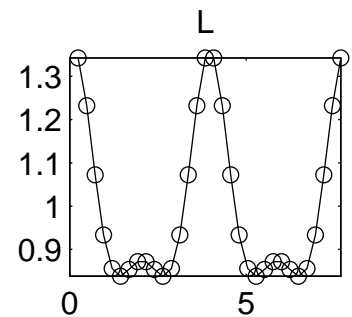
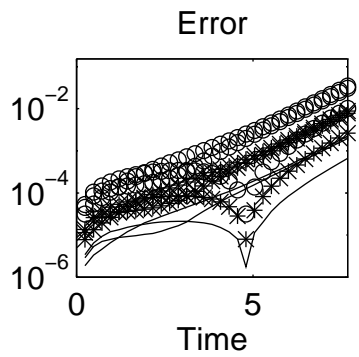
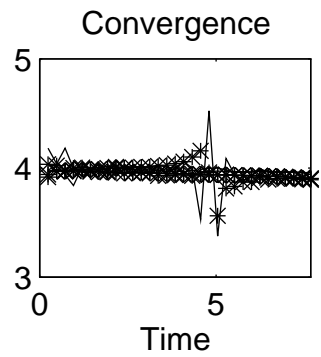
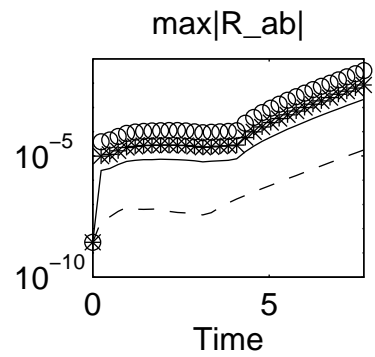
\end{document}